\documentclass[12pt]{article}
\usepackage{epsf}
\usepackage{graphicx}

\begin{document}
\title
{Asymptotic freedom in low-energy quantum gravity}
\author
{Michael A. Ivanov \\
Physics Dept.,\\
Belarus State University of Informatics and Radioelectronics, \\
6 P. Brovka Street,  BY 220027, Minsk, Republic of Belarus.\\
E-mail:  michai@mail.by.}

\maketitle

\begin{abstract} It is suggested in the model of low-energy quantum
gravity by the author, that the background of super-strong
interacting gravitons exists. It is shown here that
micro-particles at very small distances should be almost free if
the gravitational attraction is caused by a pressure of these
gravitons.
\end{abstract}
In the model of low-energy quantum gravity \cite{500}, the
Newtonian attraction arises as a result of the two stochastic
processes: a pressure of graviton pairs and a repulsion due to
single scattered gravitons after destruction of these pairs. A
force of attraction of two bodies due to pressure of graviton
pairs, $F_{2}$, is equal to (in notations of \cite{500}):
\begin{equation}
F_{2}= \int_{0}^{\infty} {\sigma (E_{2},<\epsilon_{2}>) \over 4
\pi r^{2}} \cdot 4 \sigma (E_{1},<\epsilon_{2}>)\cdot {1 \over 3}
\cdot {4 f_{2}(2\omega,T) \over c} d\omega =
\end{equation}
$$ {8 \over 3} \cdot
{D^{2} c(kT)^{6} m_{1}m_{2} \over {\pi^{3}\hbar^{3}r^{2}}}\cdot
I_{2},$$ where $I_{2}$ is a constant. Here, a portion of screened
gravitons {\it for big distances between the bodies} is described
by the factor $\sigma (E_{2},<\epsilon_{2}>) / 4 \pi r^{2},$ which
should be much smaller of unity. A net force is attractive, and it
is equal to $F_{2}/2.$ For small distances, the condition $\sigma
(E_{2},<\epsilon_{2}>) \ll 4 \pi r^{2}$ will be broken. For
example, $\sigma (E_{2},<\epsilon_{2}>) \sim 4 \pi r^{2}$ for two
protons and $<\epsilon_{2}> \sim 10^{-3} \ eV$ at distances $r
\sim 10^{-11} \ m.$ This quantity is many orders larger than the
Planck length. I noted in \cite{500} that the screened portion may
tend to a fixed value at super-short distances, and it will be
something similar to asymptotic freedom of strong interactions
\cite{20,21} - and there I stopped.
\par But when I compute a pressure force of graviton pairs in the
limit case of super-short distances I was surprised: it turns out
that this force almost vanishes. For this limit case, we should
replace the factor $\sigma (E_{2},<\epsilon_{2}>) / 4 \pi r^{2}$
by $1/2$ if a separation of interacting particles has a sense. Of
course, it is an idealization because this factor depends on a
graviton energy. If we accept this replacement, we get for the
pressure force (acting on body 1) instead of (1):
\begin{equation}
F_{2}= \int_{0}^{\infty} {1 \over 2} \cdot 4 \sigma
(E_{1},<\epsilon_{2}>)\cdot {1 \over 3} \cdot {4 f_{2}(2\omega,T)
\over c} d\omega =
\end{equation}
$$ {8 \over 3} \cdot
{D (kT)^{5} E_{1} \over {\pi^{2}\hbar^{3}c^{3}}}\cdot I_{5},$$
where $I_{5}$ is the new constant:
\begin{equation}
I_{5} \equiv \int_{0}^{\infty}{ x^{4}
(1-\exp(-(\exp(2x)-1)^{-1}))(\exp(2x)-1)^{-3} \over
\exp((\exp(2x)-1)^{-1}) \exp((\exp(x)-1)^{-1})} d x =
\end{equation}
$$4.24656 \cdot 10^{-4}.$$
\par Then the corresponding limit acceleration is equal to:
\begin{equation}
w_{lim}=  G {\pi \over { D (kT) c^{2}}}\cdot {I_{5} \over I_{2}}=
3.691 \cdot 10^{-13} \ m/s^{2},
\end{equation}
where $G$ is the Newton constant (which is computable in the
model).
\par This extremely small acceleration means that at very
small distances (which are meantime many orders of magnitude
larger than the Planck length) we have in this model the property
which never has been recognized in any model of quantum gravity:
almost full asymptotic freedom.

\end{document}